# ELECTROSTATICS OF PHASE BOUNDARIES IN COULOMB SYSTEMS

I. Iosilevskiy and A. Chigvintsev

*Moscow Institute of Physics and Technology (State University), Moscow, Russia*

Abstract. Any interface boundary in an equilibrium system of Coulomb particles is accompanied by the existence of a finite difference in the average electrostatic potential through this boundary. The discussed interface potential drop is a thermodynamic quantity. It depends on temperature only and does not depend on surface properties. The zero-temperature limit of this drop (along the coexistence curve) is an individual substance coefficient. At high temperature the drop tends to zero at critical point of gas-liquid phase transition. A special critical exponent can be defined to describe this behavior. Study of the interface potential drop is illuminative in simplified Coulomb models: i.e. for melting and evaporation in variants of One Component Plasma model (OCP), or for model of Charged Hard/Soft Spheres (CHS/CSS) etc. In all these cases properties of the potential drop can be easily calculated by the DNS-methods (direct numerical simulation) when the two–phase coexistence in Coulomb system is really simulated. Electrostatics of phase boundaries in real systems could be elucidated in analytical calculation of two-phase coexistence via finite-temperature DFT approach (density functional theory).

### General Properties

A remarkable feature of equilibrium Coulomb system is presence of non-locality features in thermodynamic equilibrium in such systems. The sequence of this non-locality is the existence of two versions of chemical potential. The ordinary chemical potential, $\mu_i(n,T)$, is presumed to be a local parameter depending on local density and temperature. In contract to this the *electro-chemical potential*, $\tilde{\mu}_i$, is essentially non-local parameter. Both the versions of chemical potential are presumed to be simply connected in uniform Coulomb systems. In this case the electro-chemical potential, $\tilde{\mu}_i$, is the sum of ordinary chemical potential, $\mu_i(n,T)$, and average electrostatic potential, $\varphi$, $\{\tilde{\mu}_i = \mu_i + Z_i e\varphi\}$. This relation is extended to the weakly non-equilibrium situations in frames of the local thermodynamic equilibrium approximation (LTE): $\tilde{\mu}_i(r) = \mu_i\{n(r), T(r)\} + Z_i e\varphi(r)$. For each charged species in a Coulomb system at equilibrium, the values of its ordinary chemical potentials in coexisting phases, $\mu_i'$ and $\mu_i''$, must not be equal under conditions of phase equilibrium. It is namely the electro-chemical potentials, $\tilde{\mu}_i$ which have the same values in coexisting phases, $(\tilde{\mu}_i)' = (\tilde{\mu}_i)''$. This equality combined with the electroneutrality condition in both phases leads to existence of the finite gap in the average electrostatic potential through the phase interface, $\Delta\varphi$.

$$\Delta\varphi \equiv \varphi''(r = +\infty) - \varphi'(r = -\infty) = [\mu_e'' - \mu_e']\, e^{-1} = [\mu_i' - \mu_i''](Ze)^{-1} \qquad (1)$$

In contrast to the work function the discussed potential drop, $\Delta\varphi = \Delta\varphi(T)$, is a thermodynamic quantity which depends on temperature only and does not depend on surface properties. The zero-temperature limit of this drop (along the gas–condensed state coexistence curve) is an individual substance thermo-electrophysical coefficient. It supplements the set of basic parameters of real material, such as sublimation energy, ionization potential, etc. It is the non-symmetry in equilibrium properties of various charged species in coexisting phases that manifest itself by the existence of the finite gap $\Delta\varphi$. It equals to zero identically for symmetrical systems like the electron-positron plasma, the restricted primitive ionic model of electrolyte solution, etc.

The potential drop, $\Delta\varphi(T)$ tends to zero at the critical point of gas-liquid phase transition. A special critical exponent can be defined to describe the behavior of $\Delta\varphi(T)$ in the vicinity of the critical point: $\Delta\varphi(T) \sim |T - T_C|^\phi$. All properties of discussed potential drop can be directly calculated by the DNS-methods (MC or MD direct numerical simulation) when both the coexisting phases in Coulomb system are explicitly simulated in combination.

ILLUSTRATIONS

Potential of Crystal-Fluid Interface in One Component Plasma Model - OCP(r)

The OCP model is studied carefully nowadays in standard version of ions (or electrons) on a rigid, (non-compressible) compensating background of opposite sign [1, 2, 7]. (Notation "r" stresses this property of background). It is well known that the only phase transition in OCP(r) – so-called Wigner crystallization – occurs in the model without any density gap ($n^*_{Fluid} \equiv n^*_{Crystal}$). Phase equilibrium condition in this case corresponds to the equality of Helmholtz free energy $F(N,V,T)$ in both coexisting phases (notation «*» below). It should be stressed that the values of the ordinary (local) chemical potential in both coexisting phases are not equal in general case. Equation (1) of present work corresponds to the statement that the double electrical layer ("surface dipole") must appear at crystal-fluid interface as a result of this inequality, so that the potential of this crystal-fluid interface should compensate exactly the mean-phase deviation, $\Delta\mu^* \equiv (\mu^*_{Crystal} - \mu^*_{Fluid})$ in the ordinary chemical potential. The values of electrochemical potential in both phases will be equal in this case

$$F(N,V,T)^*_{Crystal} = F(N,V,T)^*_{Fluid} \qquad \mu^*_{Crystal} \neq \mu^*_{Fluid} \;(!) \qquad [\tilde{\mu}_i]^*_{Crystal} = [\tilde{\mu}_i]^*_{Fluid} \qquad (2)$$

It is known that total melting curve in OCP(r) consists of three parts [3-9] (see Fig.1 in [6]):
- Low density melting of non-degenerated, classical ions (line $\Gamma = \Gamma_m = $ const.);

- High density (quantum) melting of highly degenerated ions  (line $r_S \cong (r_S)_m$ = const.);
- Transition zone between two parts, including the point of maximal melting temperature.

<u>Classical melting of Wigner crystal</u>  ($\theta^{(i)} \equiv k_B T/\varepsilon_F^{(i)} \gg 1$).   $\Gamma \equiv (4\pi n/3)^{1/3}(Ze)^2/kT \cong 178$  [7]

In this case $\mu(\Gamma_m)_{Crystal} < \mu(\Gamma_m)_{Fluid} < 0$. Therefore crystal is *positive* and fluid is *negative* in crystal-fluid interface and using known entropy change at melting [8] one has

$$Ze\Delta\varphi_{melting} = [\mu_i''(\Gamma_m) - \mu_i'(\Gamma_m)] = k_B T (\Delta S^*/3Nk_B)_{melting} \cong 0.27.. \, k_B T \qquad (3)$$

<u>"Cold" (Quantum) melting of Wigner crystal</u>  ($\theta^{(i)} \equiv k_B T/\varepsilon_F^{(i)} \ll 1$)  $\leftrightarrow$  $r_S = (r_S)_m \cong 100$  [9]

In this case one has:  $0 > \mu\{(r_S)_m\}_{Crystal} > \mu\{(r_S)_m\}_{Fluid}$, so that the fluid is *positive* and crystal is *negative* in crystal–fluid interface. Using the results of Ceperley and Alder MC-simulation [9] we estimated roughly the following value $(\Delta\mu)_{melting}$ at $T \to 0$

$$(\Delta\varphi)_{melting} = (\Delta\mu)_{melting}/(Ze) \approx -0.2 \, V \qquad (\theta^{(i)} \ll 1, T \to 0) \qquad (4)$$

<u>Intermediate melting zone and maximum melting temperature (MMT)</u>  –  ($\theta^{(i)} \sim 1$)

Estimated MMT value:  $T^{**} \equiv \max(T_{melt}) \approx (3 \div 10) \, 10^{-5}$ Ry [3-6]. In this remarkable MMT-point ($T^{**}, n^{**}$) crystal–fluid equilibrium corresponds to following special conditions:

$$(F^{**})_{Crystal} = (F^{**})_{Fluid} \qquad (P^{**})_{Crystal} = (P^{**})_{Fluid} \qquad (\mu^{**})_{Crystal} = (\mu^{**})_{Fluid} \qquad (5)$$

Thus one obtains for discussed potential drop of crystal-fluid interface in MMT-point

$$(\Delta\varphi^{**})_{melting} = 0 \qquad (6)$$

Consolidated picture of all three parts of total dependence $\Delta\varphi_{melting}(T)$ is exposed at Fig.1.

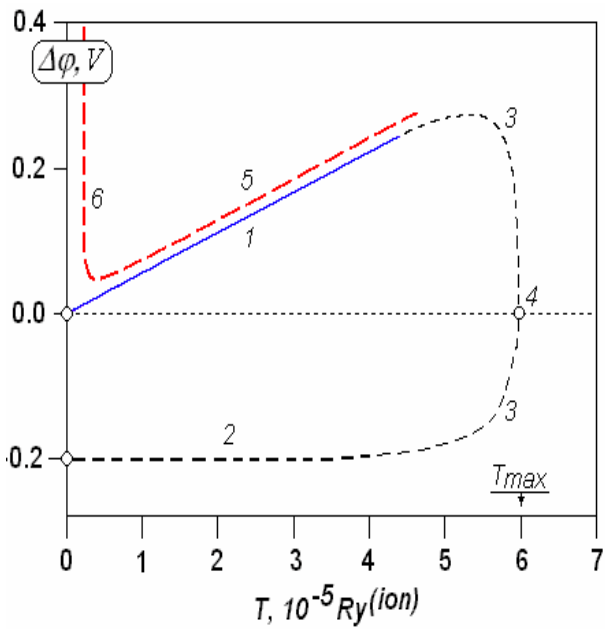

Figure 1. Potential of crystal-fluid interface in OCP($Z$ =2) {$1 \div 4$ – OCP(r); $5,6$ – OCP(c)} *1* – Melting of non-degenerated, classical ions in OCP(r)  {$\Gamma \equiv (Ze)^2/kTa \cong 178$ [7]}; *2* – Quantum melting of highly degenerated ions in OCP(r) {$r_S \equiv a/a_B \cong 100$ [9]}; *3* – Transition between *1* and *2*; *4* – $T_{max}^{**}$ – maximal melting temperature according to [6]; *5* – Melting of non-degenerated, classical ions in OCP(c) of ions ($Z = 2$) on the background of ideal gas of degenerated electrons; *6* – The same as *5* in low density limit (weakly degenerated *uniform* background). (*1, 5, 6* – present calculation [11-14]; *2* – estimation from results [9]; *3* – qualitative picture)

Electrostatics of three (not one!) phase coexistences: gas-liquid, gas-crystal and crystal-fluid in modified version of OCP(c) [8] with uniform *and* compressible background (it is stressed by notation "c") have been studied carefully in [10-14]. Results are exposed at Figure 1.

Gas-Liquid Coexistence in Simple Metal ($Z = 1$)

In this case one considers coexistence of electron-ionic system in condensed phase, $n_i' + n_e'$, with electron-ion-atomic system in vapor phase, $n_i'' + n_e'' + n_a''$. Equilibrium conditions include equality of temperature, pressure and Gibbs free energy in both phases. These equations are supplemented with equation of ionization equilibrium in vapor and with electroneutrality conditions in both phases. This set of equations fix all the concentrations: $n_i'$, $n_e'$, $n_i''$, $n_e''$, $n_a''$, and therefore, fixes uniquely the interface potential drop, $\Delta\varphi(T)$ (1)

Simple and fundamental relation for $\Delta\varphi(T)$ may be obtained in the limit $T \to 0$ (along coexistence curve). In this limit $\mu_a' \to \mu_a^0 = const$. The vapor phase is ideal. It results in: $\Delta\mu_{i,e}'' \equiv \mu_i'' - \mu_e'' \to 0$. It gives [14]:

$$e\Delta\varphi(0) = [(\Delta_s H^0 + I)/2 - \{\mu_e(0)\}_{Cond}] \qquad (7)$$

Here $\Delta_s H^0$ is the sublimation energy of metal at $T = 0$; $I$ – atomic ionization potential and $\{\mu_e(0)\}_{Cond}$ — zero-temperature limit of electronic chemical potential in condensed phase.

Acknowledgements

Present work was partially supported by Grant 2550 "Universities of Russia", Grant CRDF № MO-011-0, and by RAS Scientific Program "Physics of matter under extreme conditions.